\documentclass[twocolumn,aps,english,prl,nofootinbib,preprintnumbers]{revtex4}
\usepackage{mathrsfs}
\usepackage{hyperref}
\usepackage{graphicx}
\usepackage{amsmath, amssymb}
\usepackage{babel}
\usepackage{color}
\usepackage{slashed}

\def\beqn{\begin{eqnarray}}
\def\eeqn{\end{eqnarray}}
\def\barr{\begin{array}}
\def\earr{\end{array}}
\def\btab{\begin{tabular}}
\def\etab{\end{tabular}}
\def\bite{\begin{itemize}}
\def\eite{\end{itemize}}
\def\bcen{\begin{center}}
\def\ecen{\end{center}}

\begin{document}
\preprint{MITP/20-019}

\title{
%First lattice-pheno implications on high-precision $V_{ud}$ measurement
Joint lattice QCD - dispersion theory analysis confirms the quark-mixing top-row unitarity deficit
}

\author{Chien-Yeah Seng$^{1}$}
\author{Xu Feng$^{2,3,4,5}$}
\author{Mikhail Gorchtein$^{6,7,8}$}
%\email{gorshtey@uni-mainz.de}
\author{Lu-Chang Jin$^{9,10}$}

\affiliation{$^{1}$Helmholtz-Institut f\"{u}r Strahlen- und Kernphysik and Bethe Center for Theoretical Physics,\\
	Universit\"{a}t Bonn, 53115 Bonn, Germany}
\affiliation{$^{2}$School of Physics, Peking University, Beijing 100871, China}
\affiliation{$^{3}$Collaborative Innovation Center of Quantum Matter, Beijing 100871, China}
\affiliation{$^{4}$Center for High Energy Physics, Peking University, Beijing 100871, China}
\affiliation{$^{5}$State Key Laboratory of Nuclear Physics and Technology, Peking University, Beijing 100871, China}
\affiliation{$^{6}$Helmholtz Institute Mainz, D-55099 Mainz, Germany}
\affiliation{$^{7}$GSI Helmholtzzentrum f\"ur Schwerionenforschung, 64291 Darmstadt, Germany}
\affiliation{$^{8}$Johannes Gutenberg University, D-55099 Mainz, Germany}
\affiliation{$^{9}$RIKEN-BNL Research Center, Brookhaven National Lab, Upton, NY, 11973, USA}
\affiliation{$^{10}$Physics Department, University of Connecticut, Storrs, Connecticut 06269-3046, USA}

\date{\today}

\begin{abstract}
Recently, the first ever lattice computation of the $\gamma W$-box radiative correction to the rate of the semileptonic pion decay 
allowed for a reduction of the theory uncertainty of that rate by a factor of $\sim3$. 
A recent dispersion evaluation of the $\gamma W$-box correction on the neutron also led to a significant reduction of the theory uncertainty, but shifted the value of $V_{ud}$ extracted from the neutron and superallowed nuclear $\beta$ decay, resulting in a deficit of the CKM unitarity in the top row. A direct lattice computation of the $\gamma W$-box correction for the neutron decay would provide an independent cross-check for this result but is very challenging. Before those challenges are overcome, we propose a hybrid analysis, converting the lattice calculation on the pion to that on the neutron by a combination of dispersion theory and phenomenological input. 
The new prediction for the universal radiative correction to free and bound neutron $\beta$-decay reads $\Delta_R^V=0.02477(24)$, in excellent agreement with the dispersion theory result $\Delta_R^V=0.02467(22)$. 
%The new result for the CKM unitarity in the top row that relies on currently available calculations of nuclear structure corrections reads $|V_{ud}|^2+|V_{us}|^2+|V_{ub}|^2-1=-0.0017(6)$, and the CKM unitarity deficit persists.
Combining with other relevant information, the top-row CKM unitarity deficit persists.
%$|V_{ud}|^2+|V_{us}|^2+|V_{ub}|^2-1=-0.0017(6)$.
\end{abstract}

\maketitle

%{\bf Introduction}\\

%The good agreement \cite{Marciano:2005ec,Tanabashi:2018oca} with the first-row 

Universality of the weak interaction, conservation of vector current and completeness of the Standard Model (SM) finds its 
exact mathematical expression in the requirement of unitarity of the Cabibbo-Kobayashi-Maskawa (CKM) matrix. Of various 
combinations of the CKM matrix elements constrained by unitarity, the top-row constraint is the best known both experimentally and theoretically. The 2018 value, $\Delta^u_\mathrm{CKM}\equiv |V_{ud}|^2+|V_{us}|^2+|V_{ub}|^2-1=-0.0006(5)$ \cite{Marciano:2005ec,Tanabashi:2018oca} is in good agreement with zero required in the SM, putting severe constraints on Beyond Standard Model (BSM) physics. 

Notably, the main source of the uncertainty in the $\Delta^u_\mathrm{CKM}$ constraint is theoretical: the $\gamma W$-box radiative correction (RC), prone to effects of the strong interaction described by Quantum Chromodynamics (QCD), affects the value of $|V_{ud}|$ extracted from the free neutron and superallowed nuclear $\beta$ decays.
In a series of recent papers, this RC was reevaluated within the dispersion relation technique \cite{Seng:2018yzq,Seng:2018qru,Gorchtein:2018fxl}. In particular, Ref.\cite{Seng:2018yzq} observed  that the universal, free-neutron correction received a significant shift, later confirmed qualitatively by Ref. \cite{Czarnecki:2019mwq}. This shift is the main cause of the current apparent unitarity deficit, $\Delta_{\mathrm{CKM}}^u=-0.0016(6)$ (using an average of $V_{us}$ from $K_{\ell2}$ and $K_{\ell3}$ decays \cite{Tanabashi:2018oca}). The slight increase in the uncertainty is due to nuclear structure effects \cite{Seng:2018qru,Gorchtein:2018fxl}. 

Since in superallowed $\beta$ decays one aims for a $10^{-4}$ precision, it is highly desirable to assess the uncertainty and possible, unaccounted for, systematic effects in the non-perturbative regime of QCD in a model-independent way. 
A common limitation of the studies above is the lack of experimental data to directly constrain the hadronic matrix element relevant to the RC. By means of isospin symmetry, Ref.\cite{Seng:2018yzq} relates the input to the dispersion integral at low photon virtuality $Q^2$ to a very limited and imprecise set of data on neutrino scattering on light nuclei from the 1980s \cite{Bolognese:1982zd,Allasia:1985hw}. The analysis of Ref.\cite{Czarnecki:2019mwq} consists of pure model studies.

A complete change of landscape is expected following the first direct application of the lattice QCD to RC in leptonic meson decays, $K\to\mu\nu_\mu$ and $\pi\to\mu\nu_\mu$ \cite{Giusti:2017dwk}.
Very recently, the first ever direct lattice calculation of the RC in semi-leptonic $\beta$ decay was presented, where the relevant hadronic matrix element responsible for the $\gamma W$-box diagram in the pion is calculated to high precision as a function of $Q^2$ \cite{Feng:2020zdc}. As a result, the theory uncertainty of the $\pi_{e3}\,(\pi^-\to\pi^0e\bar{\nu}_e)$ decay rate is reduced by a factor of 3. While theoretically very clean, $\pi_{e3}$ is not the easiest avenue to extract $V_{ud}$ due to its tiny branching ratio $\sim 10^{-8}$. 
Nonetheless, it provides useful information about the involved nonperturbative dynamics, especially its low-$Q^2$ behavior and its smooth transition to the perturbative regime. Using the same method or other approaches such as Feynman-Hellmann theorem \cite{Bouchard:2016heu,Seng:2019plg}, a first-principle calculation of the RC to the free neutron $\beta$ decay, while very challenging, is expected to be performed in the near future.

In this paper, we perform a combined lattice QCD -- phenomenological analysis. Making use of a body of hadron-hadron scattering data, known meson decay widths and the guidance of Regge theory and vector dominance, along with constraints from isospin symmetry, analyticity and unitarity, we are able to unambiguously relate the input into the dispersion integral for the $\gamma W$-box RC on the pion and on the neutron. Fixing the strength of the pion matrix element from the lattice, we thus obtain an estimate of an analogous matrix element on the neutron, in accord with all the aforementioned physics constraints. 

%will demonstrate that the lattice calculation of the pion RC provides extremely powerful constraints on the low-$Q^2$ behavior of the hadronic matrix element in the neutron RC, which also enters superallowed nuclear $\beta$ decays from which $|V_{ud}|$ is extracted with highest precision. Based on that, we perform for the first time a combined lattice-phenomenology analysis on the so-called ``nucleus-independent RC" $\Delta_R^V$ in superallowed $\beta$ decays and provide a corresponding update of $V_{ud}$. Our new result is consistent with Ref.\cite{Seng:2018yzq}, but is now based on a much more realistic error analysis, where large missing systematic effects are unlikely to occur. This serves as a solid confirmation to the existing tension in the first-row CKM unitarity, which motivates immediate lattice QCD studies of the RC in neutron decay. 

\begin{figure}[t]
	\includegraphics[width=0.7\columnwidth]{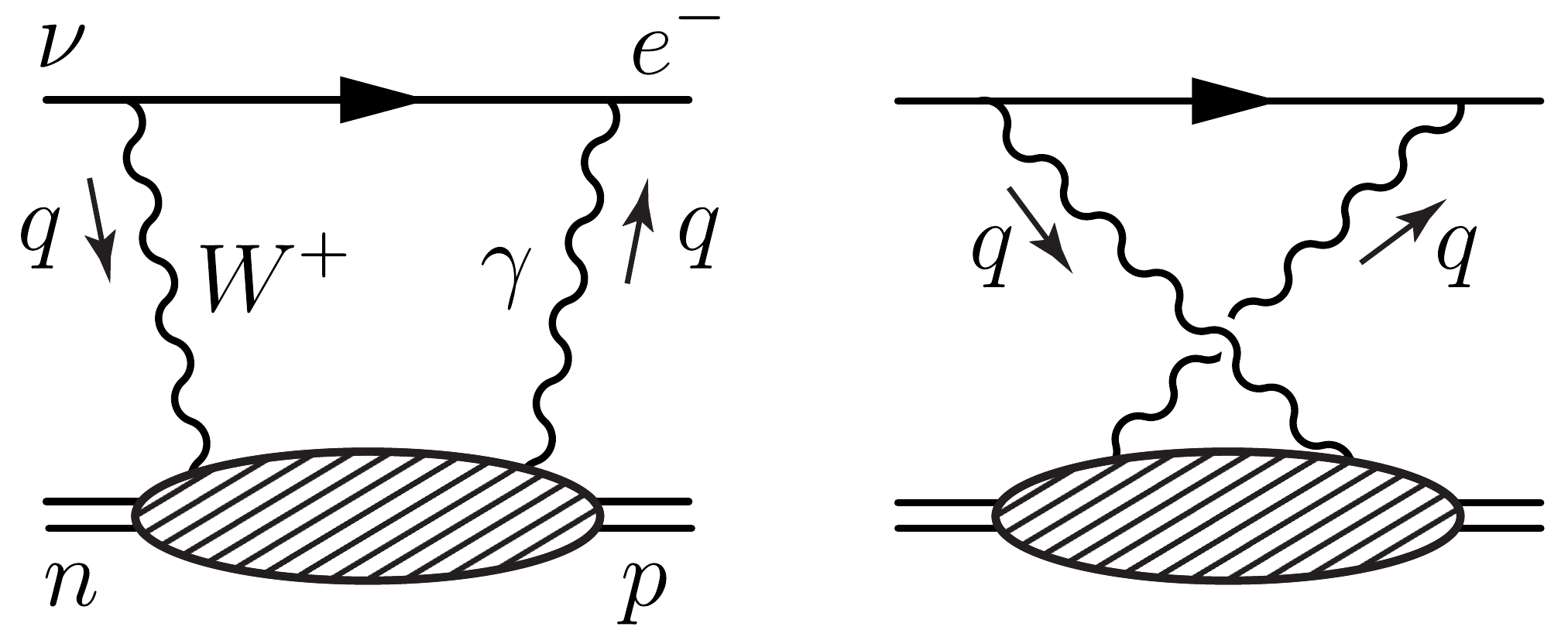}
	\caption{The $\gamma W$ box diagram in free neutron decay.}
	\label{fig:gammaWbox}
\end{figure}

We start by writing down the dispersive representation of the contribution of the $\gamma W$ box diagram (see Fig.\ref{fig:gammaWbox}) to the rate of the Fermi part of a semileptonic $\beta$ decay process of $H_i\rightarrow H_fe\bar{\nu_e}$ \cite{Seng:2018yzq,Seng:2018qru}:
\begin{equation}
\delta_{\gamma W,H}^{VA}=\frac{3\alpha}{\pi}\int_0^\infty\frac{dQ^2}{Q^2}\frac{M_W^2}{M_W^2+Q^2}M_{3H}^{(0)}(1,Q^2)
%\frac{M_{3H}^{(0)}(1,Q^2)}{F_{+}^H}
,\label{eq:deltaVA}
\end{equation}
where $\alpha$ is the fine-structure constant. 
The above definition of the $\gamma W$-box correction corresponds to a shift $|V_{ud}|^2\to |V_{ud}|^2(1+\delta_{\gamma W,H}^{VA})$, affecting the apparent value of $V_{ud}$ extracted from an experiment. 
The function
\begin{equation}
M_{3H}^{(0)}(1,Q^2)=\frac{4}{3}\int_0^1dx\frac{1+2r_H}{(1+r_H)^2}\frac{F_{3H}^{(0)}(x,Q^2)}{F_{+}^H}
%F_{3H}^{(0)}(x,Q^2)
\end{equation}
stands for the first Nachtmann moment of the (spin-independent) parity-odd structure function $F_{3H}^{(0)}(x,Q^2)$, resulting from the product between the axial charged weak current and the isoscalar electromagnetic current:
\begin{eqnarray}
&&\frac{i\epsilon^{\mu\nu\alpha\beta}p_\alpha q_\beta}{2p\cdot q}F_{3H}^{(0)}(x,Q^2)=\frac{1}{8\pi}\sum_X(2\pi)^4\delta^{(4)}(p+q-p_X)\nonumber\\
&&\times\left\langle H_f(p)\right|J_\mathrm{em}^{(0)\mu}\left|X\right\rangle\left\langle X\right|(J_W^{\nu})_A\left|H_i(p)\right\rangle.\label{eq:F30}
\end{eqnarray}
Above, $M_H$ is the average mass of $H_i,\,H_f$, $Q^2=-q^2$, $x=Q^2/2p\cdot q$,  and $r_H=\sqrt{1+4M_H^2x^2/Q^2}$, 
and the factor $F_{+}^H$ defines the normalization of the tree-level hadronic matrix element of the vector charged weak current:
\begin{equation}
\left\langle 
H_f(p)\right|\left(J_W^{\mu}\right)_V\left|H_i(p)\right\rangle=V_{ud}F_{+}^H\,2p^\mu.\label{eq:FF}
\end{equation}
By isospin symmetry, $F_+^n=1$ and $F_+^{\pi^-}=\sqrt2$. 

The quantity $\delta_{\gamma W,H}^{VA}$ is the source of the largest theory uncertainty of the RC in the $\pi_{e3}$, free neutron $\beta$ decay, and the universal RC in superallowed nuclear $\beta$ decays, and has long been the limiting factor for the precise determination of $V_{ud}$. To obtain $\delta_{\gamma W,H}^{VA}$ we need to know the Nachtmann moment $M_{3H}^{(0)}(1,Q^2)$ as a function of $Q^2$. At large $Q^2$, the product of currents in Eq. \eqref{eq:F30} is given by the leading-order (LO) operator product expansion (OPE) and the perturbative QCD (pQCD) corrections. The LO OPE + pQCD result is independent of the external state $H$ and is known up to order $\mathcal{O}(\alpha_s^4)$~\cite{Baikov:2010iw,Baikov:2010je}, with $\alpha_s$ the strong coupling constant. However, at low $Q^2$ the structure function $F_{3H}^{(0)}(x,Q^2)$ depends on details of different on-shell intermediate states $\left|X\right\rangle$ that dominate different regions of $\{x,Q^2\}$ (see Fig.2 of Ref.\cite{Seng:2018yzq} for the explanation). Also, the transition point between perturbative and non-perturbative regime is \textit{a priori} unknown, or uncertain. 

\begin{figure}[t]
	\includegraphics[width=1\columnwidth]{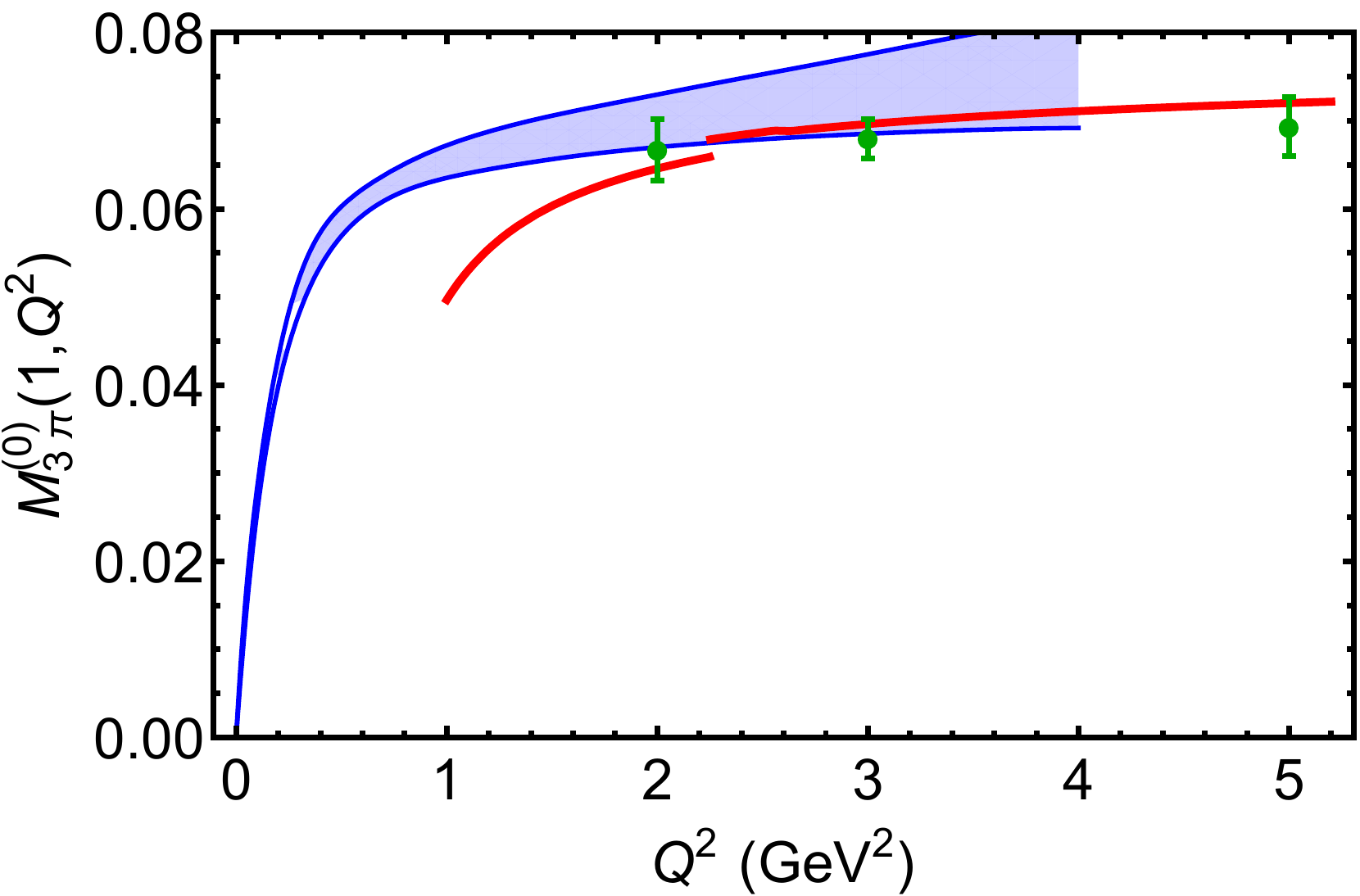}
	\caption{Comparison between the lattice calculation of $M_{3\pi}^{(0)}(1,Q^2)$ (blue band), the prediction from LO OPE with $\mathcal{O}(\alpha_s^4)$ pQCD corrections (red curve) and the low-$Q^2$ CCFR data~\cite{Kataev:1994ty,Kim:1998kia} (green points).}
	\label{fig:pionlattice}
\end{figure}

The first calculation of $M_{3\pi}^{(0)}(1,Q^2)$ on the lattice in Ref.\cite{Feng:2020zdc}
serves as an important step in addressing the questions above. Its result is presented in Fig.\ref{fig:pionlattice} as a function of $Q^2$. At low $Q^2$ where the integral \eqref{eq:deltaVA} is strongly weighted, lattice provides an extremely precise description of $M_{3\pi}^{(0)}(1,Q^2)$, but its uncertainty increases at large $Q^2$ due to the discretization error. Fortunately, at $Q^2>2\:\mathrm{GeV}^2$ there exists very precise data for the first Nachtmann moment of the parity-odd structure function $F_{3}^{\nu p+\bar{\nu}p}$ measured in the $\nu/\bar{\nu}$ scattering on light nuclei by the CCFR Collaboration~\cite{Kataev:1994ty,Kim:1998kia}. Their good agreement with pQCD prediction indicates a smooth transition to the perturbative regime at $Q^2>2\:\mathrm{GeV}^2$, which also implies that these data, upon simple rescaling, can be converted to $M_{3\pi}^{(0)}(1,Q^2)$ \footnote{Strictly speaking, the pQCD correction to $F_{3}^{\nu p+\bar{\nu}p}$ differs from that of $F_{3H}^{(0)}$ at $\mathcal{O}(\alpha_s^3)$, but such a difference is numerically insignificant at $Q^2>2\:\mathrm{GeV}^2$.}.  
On the other hand, below $2\:\mathrm{GeV}^2$ effects of generic higher-twist terms start to show up, and the LO OPE+pQCD prediction disagrees significantly with the lattice result.

%{\bf Phenomenological model}\\

We shall describe how the lattice result for $\delta_{\gamma W}^{VA}$ on the pion can be used to improve our understanding of $\delta_{\gamma W}^{VA}$ on the neutron. First, 
for the neutron we parametrized the structure function $F_{3N}^{(0)}$ (hence, also $M_{3N}^{(0)}$) as \cite{Seng:2018yzq,Seng:2018qru}:
\begin{equation}
%M_{3N}^{(0)}(1,Q^2)\approx \left(M_{3N}^{(0)}(1,Q^2)\right)_{\mathrm{el}+\mathbb{R}+\mathrm{res}+\pi N}
F_{3N}^{(0)}=F_{3N,\mathrm{el}}^{(0)}+\left\{
\begin{array}{c}
F_{3N,\mathrm{res}}^{(0)}+F_{3N,\pi N}^{(0)}+F_{3N,\mathbb{R}}^{(0)},\;Q^2\leq Q_0^2,\\
F_{3N,\mathrm{pQCD}}^{(0)},\;Q^2\geq Q_0^2,
\end{array}\right.\label{eq:F3N}
\end{equation}
where $Q_0^2\approx2$ GeV$^2$ is the scale above which the LO OPE + pQCD description is valid. 
Above, we isolated the contributions from the elastic intermediate state (el) fixed by the nucleon magnetic~\cite{Lorenz:2012tm,Lorenz:2014yda} and axial elastic form factor~\cite{Bernard:2001rs}, from the non-resonance $\pi N$ continuum ($\pi N$) in the low-energy region, from the $N^*$ resonances (res) \footnote{$\Delta$ resonances do not contribute due to the isoscalar nature of the photon.}, and the Regge contribution ($\mathbb{R}$) that allow to economically describe the multi-hadron continuum. 

In a similar way, we parametrize the pion structure function as 
\begin{equation}
%M_{3N}^{(0)}(1,Q^2)\approx \left(M_{3N}^{(0)}(1,Q^2)\right)_{\mathrm{el}+\mathbb{R}+\mathrm{res}+\pi N}
F_{3\pi}^{(0)}=\left\{
\begin{array}{c}
F_{3\pi,\mathrm{res}}^{(0)}+F_{3\pi,\mathbb{R}}^{(0)},\;Q^2\leq Q_0^2,\\
F_{3\pi,\mathrm{pQCD}}^{(0)},\;Q^2\geq Q_0^2.
\end{array}\right.\label{eq:F3pi}
\end{equation}
We note the absence of the elastic and the low-energy continuum contributions. The former is identically zero because the axial current does not couple to the spin-0 pion ground state. The latter would correspond to the non-resonant part of the $\pi\pi$ continuum in the $p$-wave; however, this partial wave is known to be entirely dominated by the $\rho^0$ resonance up to the $K\bar K$ threshold. 
%On the other hand, for the pion there is no elastic intermediate state, so the dominant contributions come from the resonance (which, from the point of view of unitarization, resums the $\pi^+\pi^-$ rescattering) and the multi-hadron intermediate states:
%\begin{equation}
%M_{3\pi}^{(0)}(1,Q^2)\approx M_{3\pi,\mathrm{res}}^{(0)}(1,Q^2)+M_{3\pi,\mathbb{R}}^{(0)}(1,Q^2).
%\end{equation}

Comparing the parameterizations of Eqs. (\ref{eq:F3N},\ref{eq:F3pi}), we make an important observation. Among the various contributions there are the process-specific ones that reside in the lower part of the spectrum (elastic, resonance and low-energy continuum). They have to be explicitly calculated for the pion and for the nucleon and cannot be related to each other. On the other hand, the asymptotic contributions (Regge and pQCD) are universal. This is the central point of our analysis. 

\begin{figure}[tb]
	\begin{centering}
		\includegraphics[width=1\columnwidth]{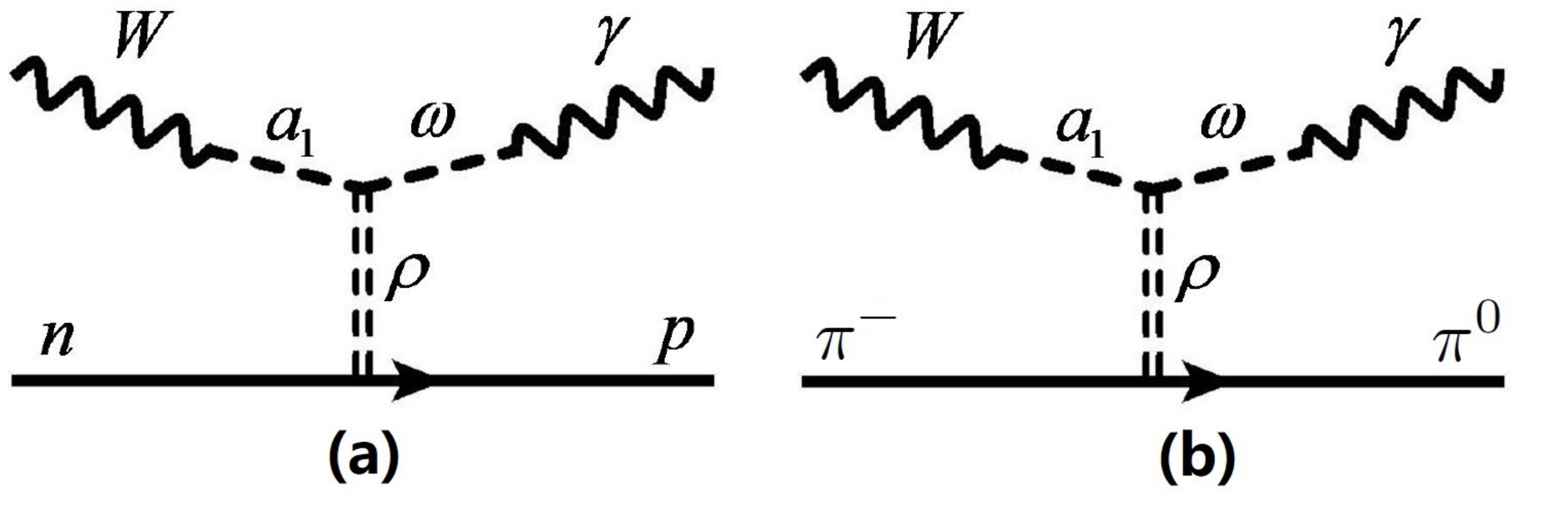}\hfill
		\par\end{centering}
	\caption{\label{fig:Regge}The Regge-exchange contribution to $F_3^{(0)}$ for neutron and pion. The vertical propagator represents the exchange of the $\rho$-trajectory. }
\end{figure}

Universality of the OPE is straightforward. The only difference between $F_{3N,\mathrm{pQCD}}^{(0)}$ and $F_{3\pi,\mathrm{pQCD}}^{(0)}$ is in the normalization of the isospin states, thus $F_{3\pi,\mathrm{pQCD}}^{(0)} = (F_+^{\pi^-}/F_+^n) F_{3N,\mathrm{pQCD}}^{(0)}$.  

 Universality is among the central predictions of Regge theory. It dictates that the upper and lower vertices in the 
Regge $\rho$-exchange amplitudes $T^{\rho}(W^++\pi^-\to\gamma+\pi^0)$ and $T^{\rho}(W^++n\to\gamma+p)$ in Fig. \ref{fig:Regge} factorize, so that, e.g.,
\beqn
R_{\pi/N}=\frac{T^{\rho}_{W^++\pi^-\to\gamma+\pi^0}}{T^{\rho}_{W^++n\to\gamma+p}}=\frac{T^{\rho}_{\pi\pi\to\pi\pi}}{T^{\rho}_{\pi N\to\pi N}}=\frac{T^{\rho}_{\pi N\to\pi N}}{T^{\rho}_{NN\to NN}},
\eeqn
where $T^{\rho}_{\pi\pi\to\pi\pi},T^{\rho}_{\pi N\to\pi N},T^{\rho}_{NN\to NN}$ stand for the amplitudes in elastic $\pi\pi,\,\pi N,\,NN$ scattering in the channel that corresponds to an exchange of the quantum numbers of the $\rho$ meson in the $t$-channel. Regge factorization has been tested on global data sets for elastic pion, pion-nucleon and nucleon-nucleon scattering. 

This leads to a prediction based on Regge universality, %{\color{red}(CYS:We say that $F_{3\pi,\mathbb{R}}$ is completely fixed by lattice. So it is more natural to put $R_{\pi/N}$ on the nucleon side.) }
\begin{eqnarray}
F_{3N,\mathbb{R}}^{(0)}(x,Q^2)&=&R_{\pi/N}^{-1}F_+^nA(Q^2)f_\mathrm{th}^N(W^2)\left(\frac{Q^2}{x}\right)^{\alpha_0^\rho}\\
F_{3\pi,\mathbb{R}}^{(0)}(x,Q^2)&=&F_+^{\pi^-} A(Q^2)f_\mathrm{th}^\pi(W^2)\left(\frac{Q^2}{x}\right)^{\alpha_0^\rho},\nonumber
\end{eqnarray}
with $\alpha_0^\rho=0.477$~\cite{Kashevarov:2017vyl}. Here we define the threshold function $f_\mathrm{th}^H=\Theta(W^2-W_{\mathrm{th},H}^2)(1-\exp[(W_{\mathrm{th},H}^2-W^2)/\Lambda^2])$, where $W^2=M_H^2+Q^2(\frac{1}{x}-1)$ and $\Lambda=1\:\mathrm{GeV}^2$ \cite{Gorchtein:2011mz}. The threshold parameter $W_{\mathrm{th},H}$ characterizes the threshold for the multi-hadron contributions. 
In Ref.~\cite{Seng:2018yzq} we fixed $W_{\mathrm{th},N}=m_N+2M_\pi$, such that the threshold function $f_\mathrm{th}^N\approx1$ for $W\gtrsim2.5$GeV. 
In the pion sector, one expects $W_{\mathrm{th},\pi}$ to lie between $M_\rho$ and $1.2$~GeV, the scale above which Regge description is valid \cite{Caprini:2011ky}. 
In this work we choose $W_{\mathrm{th},\pi}\approx1$~GeV, and account for the uncertainty due to its variation between the two boundaries. 

The function $A(Q^2)$ describes the interaction at the upper half of Fig.\ref{fig:Regge} and is, within the Regge framework, common for neutron and pion. It is generally unknown but is now completely fixed by the lattice result plotted in Fig.\ref{fig:pionlattice}---upon subtracting the resonance contribution. With these ingredients, the ratio of the first Nachtmann moments of the Regge contributions reads,
\begin{equation}
%\frac{F_+^{\pi^-}}{F_+^n}
\frac{M_{3N,\mathbb{R}}^{(0)}(1,Q^2)}{M_{3\pi,\mathbb{R}}^{(0)}(1,Q^2)}=\frac{1}{R_{\pi/N}}\frac{\int_0^1 dx\frac{1+2r_N}{(1+r_N)^2}f_\mathrm{th}^N(W^2)x^{-\alpha_0^\rho}}{\int_0^1 dx \frac{1+2r_\pi}{(1+r_\pi)^2}f_\mathrm{th}^\pi(W^2) x^{-\alpha_0^\rho}}.\label{eq:M3ratio}
\end{equation}

%Now the basic idea is that $M_{3N,\mathbb{R}}^{(0)}$ can be related to $M_{3\pi,\mathbb{R}}^{(0)}$ through the general principle of factorization, and the latter is obtainable from the lattice curve upon subtracting the resonance contribution. Below we will show how it actually works.

\begin{figure}[tb]
	\begin{centering}
		\includegraphics[width=0.6\columnwidth]{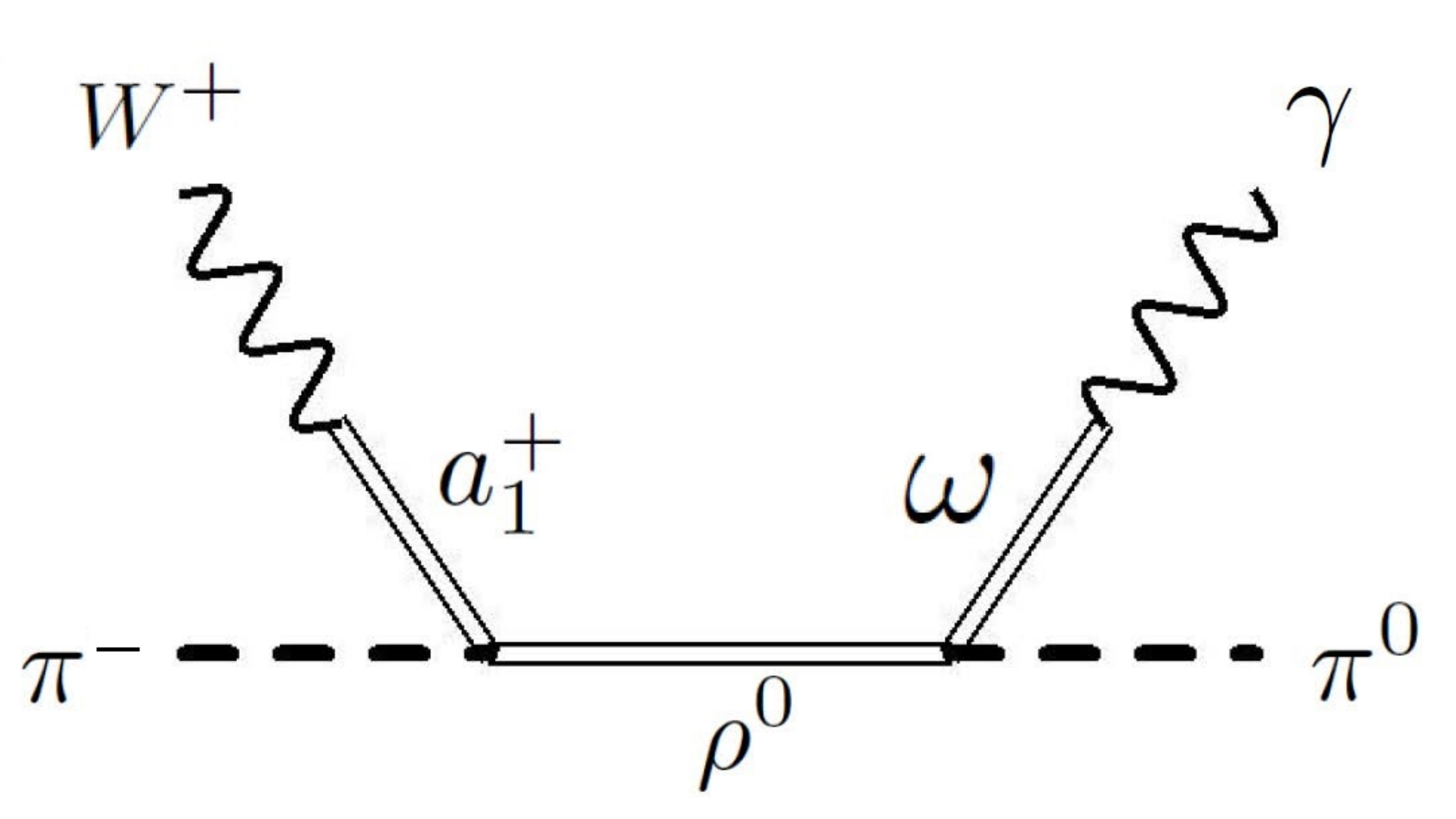}\hfill
		\par\end{centering}
	\caption{\label{fig:resonance}The $\rho$-exchange contribution to $F_{3\pi}^{(0)}$. The propagators of $\omega$ and $a_1$ mesons indicate the vector-meson-dominance form factors.}
\end{figure}

\begin{figure}[tb]
	\begin{centering}
		\includegraphics[width=1\columnwidth]{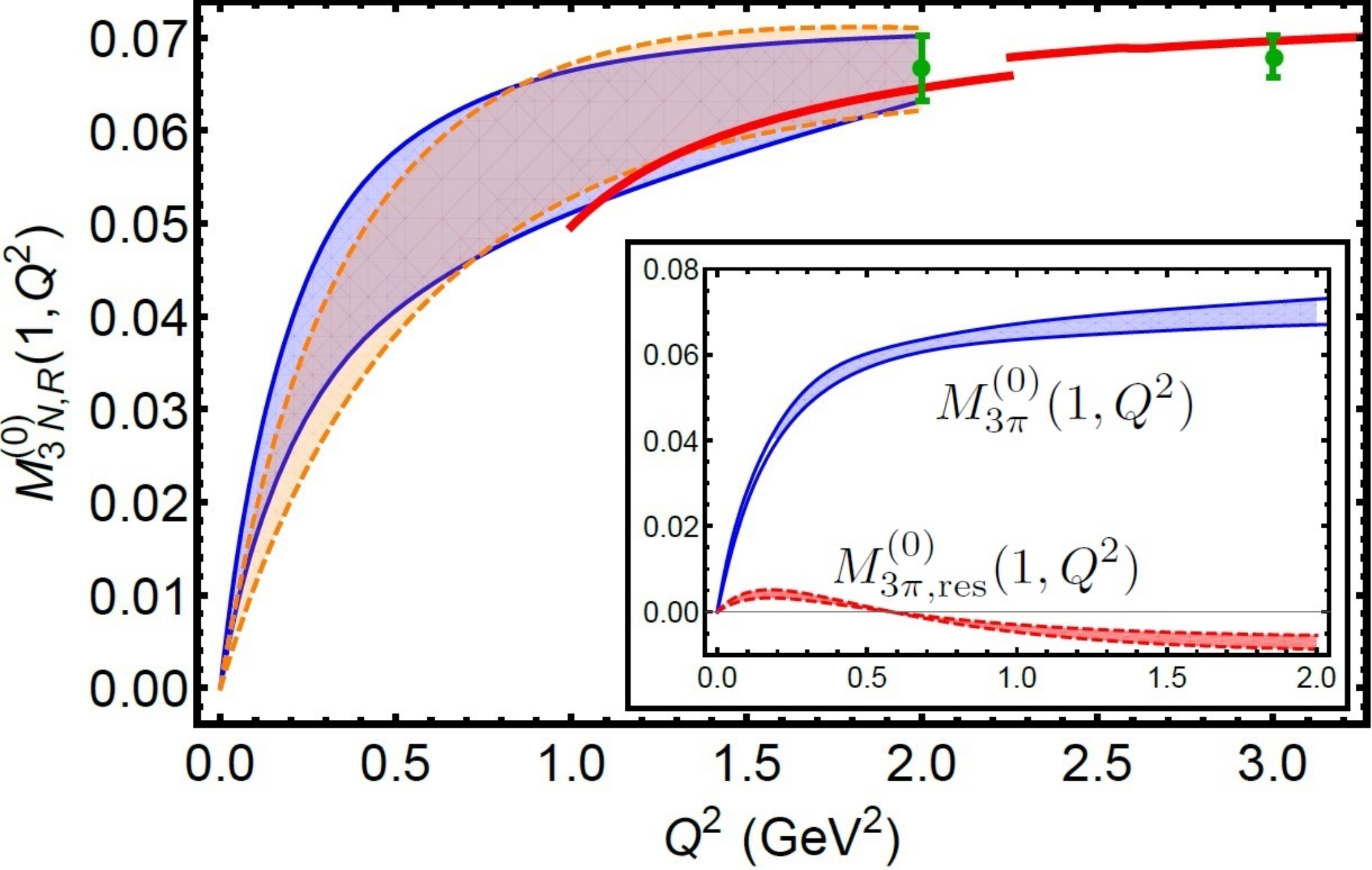}\hfill
		\par\end{centering}
	\caption{The new determination of $M_{3N,\mathbb{R}}^{(0)}(1,Q^2)$ (blue band with solid boundaries) is compared to the result of Ref.\cite{Seng:2018yzq} (orange band with dashed boundaries), the pQCD prediction (red curve) and the CCFR data~\cite{Kataev:1994ty,Kim:1998kia} (green points). In the bottom-right subview, the resonance contribution to $M_{3\pi,\,{\rm res}}^{(0)}$(red dashed curves and band) is shown along with the full lattice calculation $M_{3\pi}^{(0)}$ (blue solid curves and band).}\label{fig:old_new}
\end{figure}

To fully specify the parametrization of  $F_{3\pi}^{(0)}$ we turn now to
the resonance contribution depicted in Fig.~\ref{fig:resonance}. 
Its strength is derived from the following effective Lagrangian densities~\cite{Meissner:1987ge},
\begin{eqnarray}
\mathcal{L}_{\rho\gamma\pi}&=&\frac{eg_{\rho\gamma\pi}}{2M_\rho}
F_\omega(Q^2)
(F_\rho^a)^{\mu\nu}\tilde{F}_{\mu\nu}\pi^a\\
\mathcal{L}_{a_1\rho\pi}&=&\frac{g_{a_1\rho\pi}}{2M_{a_1}}\varepsilon^{abc}(F_\rho^a)^{\mu\nu}(F_{a_1}^b)_{\mu\nu}\pi^c\nonumber\\
\mathcal{L}_{Wa_1}&=&\frac{gM_{a_1}^2}{2g_\rho}w_{a_1}
F_{a_1}(Q^2)
V_{ud}W_\mu^- a_1^{+\mu}+h.c.,\nonumber
\end{eqnarray}
where we explicitly include the vector dominance form factors $F_{\omega, a_1}(Q^2)=[1+Q^2/M_{\omega,a_1}^2]^{-1}$. 
The couplings are obtained as follows: $|g_{\rho\gamma\pi}|=0.645(43)$ from the $\rho\rightarrow \gamma\pi$ decay width, $|g_{a_1\rho\pi}|$ is allowed to vary from 0 all the way to $5.7(1.3)$ which saturates the full $a_1$ decay width \cite{Tanabashi:2018oca}, and $|w_{a_1}/g_\rho|=0.133$ from the $\tau^-\rightarrow a_1^-\nu_\tau$ decay width\cite{Lichard:1997ya}. 
%They give, upon the narrow-width approximation of $\rho^0$, %{\color{red}(CYS:Using wide text will create too many empty space.)}
%\begin{eqnarray}
%&&%\frac{1}{F_+^{\pi^-}}
%M_{3\pi,\mathrm{res}}^{(0)}(1,Q^2)=\frac{1}{6}\frac{1+2\tilde{r}}{(1+\tilde{r})^2}\left|\frac{w_{a_1}}{g_\rho}g_{a_1\rho\pi}g_{\rho\gamma\pi}\right|\nonumber\\
%&&\times F_{a_1}(Q^2)F_\omega(Q^2)\frac{M_\rho^2-M_\pi^2-Q^2}{M_\rho^2-M_\pi^2+Q^2}\frac{Q^2}{M_{a_1}M_\rho}.\label{eq:M3Res}
%\end{eqnarray}
%\begin{widetext}
%\begin{eqnarray}
%M_{3\pi,\mathrm{res}}^{(0)}(1,Q^2)=-\frac{1}{3\sqrt{2}}\frac{1+2\tilde{r}}{(1+\tilde{r})^2}\left|\frac{w_{a_1}}{g_\rho}g_{a_1\rho\pi}g_{\rho\gamma\pi}\right|F_{a_1}(Q^2)F_\omega(Q^2)\frac{M_\rho^2-M_\pi^2-Q^2}{M_\rho^2-M_\pi^2+Q^2}\frac{Q^2}{M_{a_1}M_\rho}.
%\label{eq:res}
%\end{eqnarray}
%\end{widetext}
%where $\tilde{r}=(1+4M_\pi^2x_R^2/Q^2)^{1/2}$, $x_R=Q^2/(Q^2+M_\rho^2-M_\pi^2)$, and $F_V(Q^2)=(1+Q^2/M_V^2)^{-1}$. 
Finally, the overall sign of $M_{3\pi,\mathrm{res}}^{(0)}$ is fixed by requiring that it matches the sign of the $\pi\pi$ contribution calculated in Chiral Perturbation Theory at small $Q^2$. Numerically, the size of $M_{3\pi,\mathrm{res}}^{(0)}$ is rather small, $\leq10\%$ of the total, as can be seen in the bottom-right subview of Fig.\ref{fig:old_new} where the resonance estimate (red dashed curves and band) is plotted along with the full lattice calculation (blue curves and band). This smallness guarantees that the removal of the non-universal resonance contribution does not introduce an uncontrolled systematic uncertainty in our analysis.

With Eq.\eqref{eq:M3ratio}, $M_{3N,\mathbb{R}}^{(0)}(1,Q^2)$ could now be directly obtained from the lattice results and the rescaling factor ${R}_{\pi/N}$. A recent analysis of $\pi\pi$ scattering \cite{Caprini:2011ky} made the factorization test with respect to $\pi N$ analysis and found (omitting the isospin factor $F_+^{\pi^-}/F_+^n$),
\beqn
\frac{T^{\rho}_{\pi\pi\to\pi\pi}}{T^{\rho}_{\pi N\to\pi N}}=1.35^{+0.21}_{-0.26}.\label{eq:RpiN0}
\eeqn 
On the other hand, the OPE suggests that $R_{\pi/N}=1$ in the perturbative regime (note also the $\rho$ coupling universality hypothesis in the hidden local symmetry \cite{Sakurai:1960ju}).
Therefore, to ensure a continuous matching at all $Q^2$ values we allow $R_{\pi/N}$ to slightly depend on $Q^2$, 
\begin{equation}
R_{\pi/N}(Q^2)=R_{\pi/N}(0)+b Q^2,\label{eq:RpiNQ2}
\end{equation}
where $R_{\pi/N}(0)$ is fixed by Eq.\eqref{eq:RpiN0},
and $b$ is fixed by requiring $M_{3N,\mathbb{R}}^{(0)}$ to reproduce the CCFR datum at the matching point $Q_0^2=2$ GeV$^2$,
	\begin{equation}
	M_{3N,\mathbb{R}}^{(0)}(1,Q_0^2)=0.0667(35).
	\end{equation}
The result reads $b=-0.076^{+0.100}_{-0.072}\:\mathrm{GeV}^{-2}$.

With the prescription above we fully fix $M_{3N,\mathbb{R}}^{(0)}(1,Q^2)$ at low $Q^2$ using the lattice curve of $M_{3\pi}^{(0)}(1,Q^2)$. The result is shown in Fig.\ref{fig:old_new}, with the uncertainties from $R_{\pi/N}(Q^2)$ and $W_{\mathrm{th},\pi}$ added in quadrature. 
Integrating over $Q^2$ gives an updated estimate of the Regge contribution to $\delta_{\gamma W,N}^{VA}$:
\begin{equation}
\left(\delta_{\gamma W,N}^{VA}\right)_{\mathbb{R}}=1.12(16)_a(9)_b(3)_c\times 10^{-3}\label{eq:deltaRchange},
\end{equation}
where the uncertainties are from (a) the pion-nucleon matching, including the rescale factor $R_{\pi/N}$ and the lattice uncertainty, (b) the Regge parameterization and (c) the resonance subtraction. Our result is 
in excellent agreement with the previous determination $\left(\delta_{\gamma W,N}^{VA}\right)_{\mathbb{R}}=1.02(16)\times 10^{-3}$ \cite{Seng:2018yzq}. 
One can also study the effect of varying the perturbative matching point by evaluating the $Q^2$-integral in Eq.~\eqref{eq:deltaVA} between $2$~GeV$^2$ and $3$~GeV$^2$ using the CCFR data instead of the pQCD expression. That gives an insignificant extra uncertainty of $1\times 10^{-5}$, confirming the robustness of our error analysis. 

We next discuss the impact of this result on the extraction of $V_{ud}$. From superallowed nuclear $\beta$ decay, we have \cite{Hardy:2014qxa}:
\begin{equation}
|V_{ud}|^2=\frac{2984.43\:\mathrm{s}}{\mathcal{F}t(1+\Delta_R^V)},\:\:(\mathrm{superallowed})
\end{equation}
where $\mathcal{F}t$ is the $ft$-value corrected by nuclear effects, $\Delta_R^V=\delta_{\gamma W,N}^{VA}+...$ is the nucleus-independent RC that contains the largest theoretical error. In this paper we update the Regge contribution to $\delta_{\gamma W,N}^{VA}$ according to Eq.\eqref{eq:deltaRchange}. Meanwhile, we also update the pQCD contribution above $2\:\mathrm{GeV}^2$ from $\mathcal{O}(\alpha_s^3)$ to $\mathcal{O}(\alpha_s^4)$\cite{Baikov:2010je,Baikov:2010iw}, which reduces $\Delta_R^V$ by mere $1\times10^{-5}$. As a result we obtain a slight shift upward with respect to the result of Ref.\cite{Seng:2018yzq}:
\begin{equation}
\Delta_R^V=0.02467(22)\rightarrow0.02477(24).\label{eq:DeltaVRshift}
\end{equation}
The recent Ref. \cite{Czarnecki:2019mwq} estimated a lower value, $\Delta_R^V=0.02426(32)$, based on the assumption that the full Nachtmann moment should follow the perturbative curve down to as far as $Q^2=1$ GeV$^2$, and only afterwards higher-twist effects (estimated in a holographic QCD model) become important. The lattice calculation on the pion \cite{Feng:2020zdc} suggests that already at $Q^2\leq2$ GeV$^2$ the higher twist contributions are non-negligible. 

The implication of Eq.\eqref{eq:DeltaVRshift} on $V_{ud}$ is as follows. First, if we take $\mathcal{F}t=3072.07(63)\:\mathrm{s}$ \cite{Hardy:2018zsb}, then $|V_{ud}|=0.97365(15)$. However, recent studies in Ref.\cite{Seng:2018qru,Gorchtein:2018fxl} unveil two mutually competing new nuclear corrections (NNC) whose net effect is to enhance the uncertainty, $\mathcal{F}t=3072(2)\:\mathrm{s}.$
Taking that into account gives $|V_{ud}|=0.97366(33)$. 
For completeness, we also quote the impact of our result to neutron beta decay, where $V_{ud}$ is determined by~\cite{Czarnecki:2004cw}: 
\begin{equation}
|V_{ud}|^2=\frac{5099.34\:\mathrm{s}}{\tau_n(1+3\lambda^2)(1+\Delta_R)}.\:\:(\mathrm{neutron})
\end{equation}
Our new analysis implies $\Delta_R=0.04002(24)$ ($\Delta_R$ is the sum of $\Delta_R^V$ and the Sirlin's function~\cite{Sirlin:1967zza}), which leads to $|V_{ud}|=0.97297(58)$ given the neutron lifetime $\tau_n=879.7(8)$s \cite{Serebrov:2017bzo,Pattie:2017vsj,Ezhov:2014tna} and the axial-vector ratio  $\lambda=-1.27641(56)$\cite{Markisch:2018ndu,Chang:2018uxx}. The result is consistent with that from the superallowed nuclear $\beta$ decays. 

Finally, we discuss the current situation of the top-row CKM unitarity. There are two different measurements of $V_{us}$, using $K_{\ell2}$ \cite{Tanabashi:2018oca} and $K_{\ell3}$ \cite{Bazavov:2018kjg} decay separately:
\begin{equation}
|V_{us}|_{K_{\ell2}}=0.2253(7),\:\:|V_{us}|_{K_{\ell3}}=0.2233(6).
\end{equation}
They disagree with each other at 2$\sigma$ level, $K_{\ell3}$ giving a smaller $|V_{us}|$ which leads to a larger unitarity violation. This, however, depends critically on the existing lattice calculation of the $K\pi$ vector form factor $f_+^{K^0\pi^-}(0)$ which is recently questioned by theory \cite{Czarnecki:2019iwz} and a new lattice paper \cite{Kakazu:2019ltq}. Another possible issue is the electromagnetic RC in $K_{\ell3}$, which may be re-analyzed in a dispersive approach~\cite{Seng:2019lxf}. We summarize the resulting $\Delta^u_\mathrm{CKM}$ from different combinations in Table \ref{tab:DeltaCKM}. In short, we observe a $(3-5)\sigma$ unitarity violation excluding the NNC, and $(1.7-3)\sigma$ violation with the NNC. Our results can be tested with a future, direct lattice calculation of the $\gamma W$-box on the neutron. After that, the emphasis should be shifted to a reassessment of the nuclear structure corrections that enter the analysis of superallowed nuclear decay. 

\begin{table}
	\begin{centering}
		\begin{tabular}{|c|c|c|c|}
			\hline 
			& $|V_{ud}|$ & $\Delta_{\mathrm{CKM}}^u$with $K_{\ell2}$ & $\Delta_{\mathrm{CKM}}^u$with $K_{\ell3}$\tabularnewline
			\hline 
			\hline 
			w/o NNC & 0.97365(15) & -0.0012(4) & -0.0021(4)\tabularnewline
			\hline 
			w/ NNC & 0.97366(33) & -0.0012(7) & -0.0021(7)\tabularnewline
			\hline 
		\end{tabular}
		\par\end{centering}
	\caption{Summary of $\Delta_{\mathrm{CKM}}^u$ for different cases.\label{tab:DeltaCKM}}
	
\end{table}
 
\begin{acknowledgments}
We appreciate Guido Martinelli and Ulf-G. Mei{\ss}ner for inspiring discussions. The work of C.Y.S. is supported in part by the DFG 
(Grant No. TRR110) and the NSFC (Grant No. 11621131001) through the funds 
provided to the Sino-German CRC 110 ``Symmetries and the Emergence of Structure in QCD'', 
and also by the Alexander von Humboldt Foundation through the Humboldt Research Fellowship. 
M.G.  is supported by EU Horizon 2020 research and innovation programme, STRONG-2020 project, under grant agreement No 824093 and  by the German-Mexican research collaboration Grant No. 278017 (CONACyT) and No. SP 778/4-1 (DFG).
X.F. was supported in part by NSFC of China under Grant No. 11775002.
L.C.J. acknowledges support by DOE grant DE-SC0010339. 
\end{acknowledgments}

\bibliography{pion_ref}

\begin{thebibliography}{38}
\expandafter\ifx\csname natexlab\endcsname\relax\def\natexlab#1{#1}\fi
\expandafter\ifx\csname bibnamefont\endcsname\relax
  \def\bibnamefont#1{#1}\fi
\expandafter\ifx\csname bibfnamefont\endcsname\relax
  \def\bibfnamefont#1{#1}\fi
\expandafter\ifx\csname citenamefont\endcsname\relax
  \def\citenamefont#1{#1}\fi
\expandafter\ifx\csname url\endcsname\relax
  \def\url#1{\texttt{#1}}\fi
\expandafter\ifx\csname urlprefix\endcsname\relax\def\urlprefix{URL }\fi
\providecommand{\bibinfo}[2]{#2}
\providecommand{\eprint}[2][]{\url{#2}}

\bibitem[{\citenamefont{Marciano and Sirlin}(2006)}]{Marciano:2005ec}
\bibinfo{author}{\bibfnamefont{W.~J.} \bibnamefont{Marciano}} \bibnamefont{and}
  \bibinfo{author}{\bibfnamefont{A.}~\bibnamefont{Sirlin}},
  \bibinfo{journal}{Phys. Rev. Lett.} \textbf{\bibinfo{volume}{96}},
  \bibinfo{pages}{032002} (\bibinfo{year}{2006}), \eprint{hep-ph/0510099}.

\bibitem[{\citenamefont{Tanabashi et~al.}(2018)}]{Tanabashi:2018oca}
\bibinfo{author}{\bibfnamefont{M.}~\bibnamefont{Tanabashi}}
  \bibnamefont{et~al.} (\bibinfo{collaboration}{Particle Data Group}),
  \bibinfo{journal}{Phys. Rev.} \textbf{\bibinfo{volume}{D98}},
  \bibinfo{pages}{030001} (\bibinfo{year}{2018}).

\bibitem[{\citenamefont{Seng et~al.}(2018)\citenamefont{Seng, Gorchtein, Patel,
  and Ramsey-Musolf}}]{Seng:2018yzq}
\bibinfo{author}{\bibfnamefont{C.-Y.} \bibnamefont{Seng}},
  \bibinfo{author}{\bibfnamefont{M.}~\bibnamefont{Gorchtein}},
  \bibinfo{author}{\bibfnamefont{H.~H.} \bibnamefont{Patel}}, \bibnamefont{and}
  \bibinfo{author}{\bibfnamefont{M.~J.} \bibnamefont{Ramsey-Musolf}},
  \bibinfo{journal}{Phys. Rev. Lett.} \textbf{\bibinfo{volume}{121}},
  \bibinfo{pages}{241804} (\bibinfo{year}{2018}), \eprint{1807.10197}.

\bibitem[{\citenamefont{Seng et~al.}(2019)\citenamefont{Seng, Gorchtein, and
  Ramsey-Musolf}}]{Seng:2018qru}
\bibinfo{author}{\bibfnamefont{C.~Y.} \bibnamefont{Seng}},
  \bibinfo{author}{\bibfnamefont{M.}~\bibnamefont{Gorchtein}},
  \bibnamefont{and} \bibinfo{author}{\bibfnamefont{M.~J.}
  \bibnamefont{Ramsey-Musolf}}, \bibinfo{journal}{Phys. Rev.}
  \textbf{\bibinfo{volume}{D100}}, \bibinfo{pages}{013001}
  (\bibinfo{year}{2019}), \eprint{1812.03352}.

\bibitem[{\citenamefont{Gorchtein}(2019)}]{Gorchtein:2018fxl}
\bibinfo{author}{\bibfnamefont{M.}~\bibnamefont{Gorchtein}},
  \bibinfo{journal}{Phys. Rev. Lett.} \textbf{\bibinfo{volume}{123}},
  \bibinfo{pages}{042503} (\bibinfo{year}{2019}), \eprint{1812.04229}.

\bibitem[{\citenamefont{Czarnecki et~al.}(2019)\citenamefont{Czarnecki,
  Marciano, and Sirlin}}]{Czarnecki:2019mwq}
\bibinfo{author}{\bibfnamefont{A.}~\bibnamefont{Czarnecki}},
  \bibinfo{author}{\bibfnamefont{W.~J.} \bibnamefont{Marciano}},
  \bibnamefont{and} \bibinfo{author}{\bibfnamefont{A.}~\bibnamefont{Sirlin}},
  \bibinfo{journal}{Phys. Rev.} \textbf{\bibinfo{volume}{D100}},
  \bibinfo{pages}{073008} (\bibinfo{year}{2019}), \eprint{1907.06737}.

\bibitem[{\citenamefont{Bolognese et~al.}(1983)\citenamefont{Bolognese, Fritze,
  Morfin, Perkins, Powell, and Scott}}]{Bolognese:1982zd}
\bibinfo{author}{\bibfnamefont{T.}~\bibnamefont{Bolognese}},
  \bibinfo{author}{\bibfnamefont{P.}~\bibnamefont{Fritze}},
  \bibinfo{author}{\bibfnamefont{J.}~\bibnamefont{Morfin}},
  \bibinfo{author}{\bibfnamefont{D.~H.} \bibnamefont{Perkins}},
  \bibinfo{author}{\bibfnamefont{K.}~\bibnamefont{Powell}}, \bibnamefont{and}
  \bibinfo{author}{\bibfnamefont{W.~G.} \bibnamefont{Scott}}
  (\bibinfo{collaboration}{Aachen-Bonn-CERN-Democritos-London-Oxford-Saclay}),
  \bibinfo{journal}{Phys. Rev. Lett.} \textbf{\bibinfo{volume}{50}},
  \bibinfo{pages}{224} (\bibinfo{year}{1983}).

\bibitem[{\citenamefont{Allasia et~al.}(1985)}]{Allasia:1985hw}
\bibinfo{author}{\bibfnamefont{D.}~\bibnamefont{Allasia}} \bibnamefont{et~al.},
  \bibinfo{journal}{Z. Phys.} \textbf{\bibinfo{volume}{C28}},
  \bibinfo{pages}{321} (\bibinfo{year}{1985}).

\bibitem[{\citenamefont{Giusti et~al.}(2018)\citenamefont{Giusti, Lubicz,
  Martinelli, Sachrajda, Sanfilippo, Simula, Tantalo, and
  Tarantino}}]{Giusti:2017dwk}
\bibinfo{author}{\bibfnamefont{D.}~\bibnamefont{Giusti}},
  \bibinfo{author}{\bibfnamefont{V.}~\bibnamefont{Lubicz}},
  \bibinfo{author}{\bibfnamefont{G.}~\bibnamefont{Martinelli}},
  \bibinfo{author}{\bibfnamefont{C.~T.} \bibnamefont{Sachrajda}},
  \bibinfo{author}{\bibfnamefont{F.}~\bibnamefont{Sanfilippo}},
  \bibinfo{author}{\bibfnamefont{S.}~\bibnamefont{Simula}},
  \bibinfo{author}{\bibfnamefont{N.}~\bibnamefont{Tantalo}}, \bibnamefont{and}
  \bibinfo{author}{\bibfnamefont{C.}~\bibnamefont{Tarantino}},
  \bibinfo{journal}{Phys. Rev. Lett.} \textbf{\bibinfo{volume}{120}},
  \bibinfo{pages}{072001} (\bibinfo{year}{2018}), \eprint{1711.06537}.

\bibitem[{\citenamefont{Feng et~al.}(2020)\citenamefont{Feng, Gorchtein, Jin,
  Ma, and Seng}}]{Feng:2020zdc}
\bibinfo{author}{\bibfnamefont{X.}~\bibnamefont{Feng}},
  \bibinfo{author}{\bibfnamefont{M.}~\bibnamefont{Gorchtein}},
  \bibinfo{author}{\bibfnamefont{L.-C.} \bibnamefont{Jin}},
  \bibinfo{author}{\bibfnamefont{P.-X.} \bibnamefont{Ma}}, \bibnamefont{and}
  \bibinfo{author}{\bibfnamefont{C.-Y.} \bibnamefont{Seng}},
  \bibinfo{journal}{Phys. Rev. Lett.} \textbf{\bibinfo{volume}{124}},
  \bibinfo{pages}{192002} (\bibinfo{year}{2020}), \eprint{2003.09798}.

\bibitem[{\citenamefont{Bouchard et~al.}(2017)\citenamefont{Bouchard, Chang,
  Kurth, Orginos, and Walker-Loud}}]{Bouchard:2016heu}
\bibinfo{author}{\bibfnamefont{C.}~\bibnamefont{Bouchard}},
  \bibinfo{author}{\bibfnamefont{C.~C.} \bibnamefont{Chang}},
  \bibinfo{author}{\bibfnamefont{T.}~\bibnamefont{Kurth}},
  \bibinfo{author}{\bibfnamefont{K.}~\bibnamefont{Orginos}}, \bibnamefont{and}
  \bibinfo{author}{\bibfnamefont{A.}~\bibnamefont{Walker-Loud}},
  \bibinfo{journal}{Phys. Rev.} \textbf{\bibinfo{volume}{D96}},
  \bibinfo{pages}{014504} (\bibinfo{year}{2017}), \eprint{1612.06963}.

\bibitem[{\citenamefont{Seng and Mei{\ss}ner}(2019)}]{Seng:2019plg}
\bibinfo{author}{\bibfnamefont{C.-Y.} \bibnamefont{Seng}} \bibnamefont{and}
  \bibinfo{author}{\bibfnamefont{U.-G.} \bibnamefont{Mei{\ss}ner}},
  \bibinfo{journal}{Phys. Rev. Lett.} \textbf{\bibinfo{volume}{122}},
  \bibinfo{pages}{211802} (\bibinfo{year}{2019}), \eprint{1903.07969}.

\bibitem[{\citenamefont{Baikov et~al.}(2010{\natexlab{a}})\citenamefont{Baikov,
  Chetyrkin, and Kuhn}}]{Baikov:2010iw}
\bibinfo{author}{\bibfnamefont{P.~A.} \bibnamefont{Baikov}},
  \bibinfo{author}{\bibfnamefont{K.~G.} \bibnamefont{Chetyrkin}},
  \bibnamefont{and} \bibinfo{author}{\bibfnamefont{J.~H.} \bibnamefont{Kuhn}},
  \bibinfo{journal}{Nucl. Phys. Proc. Suppl.}
  \textbf{\bibinfo{volume}{205-206}}, \bibinfo{pages}{237}
  (\bibinfo{year}{2010}{\natexlab{a}}), \eprint{1007.0478}.

\bibitem[{\citenamefont{Baikov et~al.}(2010{\natexlab{b}})\citenamefont{Baikov,
  Chetyrkin, and Kuhn}}]{Baikov:2010je}
\bibinfo{author}{\bibfnamefont{P.~A.} \bibnamefont{Baikov}},
  \bibinfo{author}{\bibfnamefont{K.~G.} \bibnamefont{Chetyrkin}},
  \bibnamefont{and} \bibinfo{author}{\bibfnamefont{J.~H.} \bibnamefont{Kuhn}},
  \bibinfo{journal}{Phys. Rev. Lett.} \textbf{\bibinfo{volume}{104}},
  \bibinfo{pages}{132004} (\bibinfo{year}{2010}{\natexlab{b}}),
  \eprint{1001.3606}.

\bibitem[{\citenamefont{Kataev and Sidorov}(1994)}]{Kataev:1994ty}
\bibinfo{author}{\bibfnamefont{A.~L.} \bibnamefont{Kataev}} \bibnamefont{and}
  \bibinfo{author}{\bibfnamefont{A.~V.} \bibnamefont{Sidorov}}, in
  \emph{\bibinfo{booktitle}{{'94 QCD and high-energy hadronic interactions.
  Proceedings, Hadronic Session of the 29th Rencontres de Moriond, Moriond
  Particle Physics Meeting, Meribel les Allues, France, March 19-26, 1994}}}
  (\bibinfo{year}{1994}), pp. \bibinfo{pages}{189--198},
  \eprint{hep-ph/9405254}.

\bibitem[{\citenamefont{Kim et~al.}(1998)}]{Kim:1998kia}
\bibinfo{author}{\bibfnamefont{J.~H.} \bibnamefont{Kim}} \bibnamefont{et~al.},
  \bibinfo{journal}{Phys. Rev. Lett.} \textbf{\bibinfo{volume}{81}},
  \bibinfo{pages}{3595} (\bibinfo{year}{1998}), \eprint{hep-ex/9808015}.

\bibitem[{\citenamefont{Lorenz et~al.}(2012)\citenamefont{Lorenz, Hammer, and
  Mei{\ss}ner}}]{Lorenz:2012tm}
\bibinfo{author}{\bibfnamefont{I.~T.} \bibnamefont{Lorenz}},
  \bibinfo{author}{\bibfnamefont{H.~W.} \bibnamefont{Hammer}},
  \bibnamefont{and} \bibinfo{author}{\bibfnamefont{U.-G.}
  \bibnamefont{Mei{\ss}ner}}, \bibinfo{journal}{Eur. Phys. J.}
  \textbf{\bibinfo{volume}{A48}}, \bibinfo{pages}{151} (\bibinfo{year}{2012}),
  \eprint{1205.6628}.

\bibitem[{\citenamefont{Lorenz et~al.}(2015)\citenamefont{Lorenz, Mei{\ss}ner,
  Hammer, and Dong}}]{Lorenz:2014yda}
\bibinfo{author}{\bibfnamefont{I.~T.} \bibnamefont{Lorenz}},
  \bibinfo{author}{\bibfnamefont{U.-G.} \bibnamefont{Mei{\ss}ner}},
  \bibinfo{author}{\bibfnamefont{H.~W.} \bibnamefont{Hammer}},
  \bibnamefont{and} \bibinfo{author}{\bibfnamefont{Y.~B.} \bibnamefont{Dong}},
  \bibinfo{journal}{Phys. Rev.} \textbf{\bibinfo{volume}{D91}},
  \bibinfo{pages}{014023} (\bibinfo{year}{2015}), \eprint{1411.1704}.

\bibitem[{\citenamefont{Bernard et~al.}(2002)\citenamefont{Bernard,
  Elouadrhiri, and Mei{\ss}ner}}]{Bernard:2001rs}
\bibinfo{author}{\bibfnamefont{V.}~\bibnamefont{Bernard}},
  \bibinfo{author}{\bibfnamefont{L.}~\bibnamefont{Elouadrhiri}},
  \bibnamefont{and} \bibinfo{author}{\bibfnamefont{U.-G.}
  \bibnamefont{Mei{\ss}ner}}, \bibinfo{journal}{J. Phys.}
  \textbf{\bibinfo{volume}{G28}}, \bibinfo{pages}{R1} (\bibinfo{year}{2002}),
  \eprint{hep-ph/0107088}.

\bibitem[{\citenamefont{Kashevarov et~al.}(2017)\citenamefont{Kashevarov,
  Ostrick, and Tiator}}]{Kashevarov:2017vyl}
\bibinfo{author}{\bibfnamefont{V.~L.} \bibnamefont{Kashevarov}},
  \bibinfo{author}{\bibfnamefont{M.}~\bibnamefont{Ostrick}}, \bibnamefont{and}
  \bibinfo{author}{\bibfnamefont{L.}~\bibnamefont{Tiator}},
  \bibinfo{journal}{Phys. Rev.} \textbf{\bibinfo{volume}{C96}},
  \bibinfo{pages}{035207} (\bibinfo{year}{2017}), \eprint{1706.07376}.

\bibitem[{\citenamefont{Gorchtein et~al.}(2011)\citenamefont{Gorchtein,
  Horowitz, and Ramsey-Musolf}}]{Gorchtein:2011mz}
\bibinfo{author}{\bibfnamefont{M.}~\bibnamefont{Gorchtein}},
  \bibinfo{author}{\bibfnamefont{C.~J.} \bibnamefont{Horowitz}},
  \bibnamefont{and} \bibinfo{author}{\bibfnamefont{M.~J.}
  \bibnamefont{Ramsey-Musolf}}, \bibinfo{journal}{Phys. Rev.}
  \textbf{\bibinfo{volume}{C84}}, \bibinfo{pages}{015502}
  (\bibinfo{year}{2011}), \eprint{1102.3910}.

\bibitem[{\citenamefont{Caprini et~al.}(2012)\citenamefont{Caprini, Colangelo,
  and Leutwyler}}]{Caprini:2011ky}
\bibinfo{author}{\bibfnamefont{I.}~\bibnamefont{Caprini}},
  \bibinfo{author}{\bibfnamefont{G.}~\bibnamefont{Colangelo}},
  \bibnamefont{and}
  \bibinfo{author}{\bibfnamefont{H.}~\bibnamefont{Leutwyler}},
  \bibinfo{journal}{Eur. Phys. J.} \textbf{\bibinfo{volume}{C72}},
  \bibinfo{pages}{1860} (\bibinfo{year}{2012}), \eprint{1111.7160}.

\bibitem[{\citenamefont{Mei{\ss}ner}(1988)}]{Meissner:1987ge}
\bibinfo{author}{\bibfnamefont{U.~G.} \bibnamefont{Mei{\ss}ner}},
  \bibinfo{journal}{Phys. Rept.} \textbf{\bibinfo{volume}{161}},
  \bibinfo{pages}{213} (\bibinfo{year}{1988}).

\bibitem[{\citenamefont{Lichard}(1997)}]{Lichard:1997ya}
\bibinfo{author}{\bibfnamefont{P.}~\bibnamefont{Lichard}},
  \bibinfo{journal}{Phys. Rev.} \textbf{\bibinfo{volume}{D55}},
  \bibinfo{pages}{5385} (\bibinfo{year}{1997}), \eprint{hep-ph/9702345}.

\bibitem[{\citenamefont{Sakurai}(1960)}]{Sakurai:1960ju}
\bibinfo{author}{\bibfnamefont{J.~J.} \bibnamefont{Sakurai}},
  \bibinfo{journal}{Annals Phys.} \textbf{\bibinfo{volume}{11}},
  \bibinfo{pages}{1} (\bibinfo{year}{1960}).

\bibitem[{\citenamefont{Hardy and Towner}(2015)}]{Hardy:2014qxa}
\bibinfo{author}{\bibfnamefont{J.~C.} \bibnamefont{Hardy}} \bibnamefont{and}
  \bibinfo{author}{\bibfnamefont{I.~S.} \bibnamefont{Towner}},
  \bibinfo{journal}{Phys. Rev.} \textbf{\bibinfo{volume}{C91}},
  \bibinfo{pages}{025501} (\bibinfo{year}{2015}), \eprint{1411.5987}.

\bibitem[{\citenamefont{Hardy and Towner}(2018)}]{Hardy:2018zsb}
\bibinfo{author}{\bibfnamefont{J.~C.} \bibnamefont{Hardy}} \bibnamefont{and}
  \bibinfo{author}{\bibfnamefont{I.~S.} \bibnamefont{Towner}}, in
  \emph{\bibinfo{booktitle}{{13th Conference on the Intersections of Particle
  and Nuclear Physics (CIPANP 2018) Palm Springs, California, USA, May 29-June
  3, 2018}}} (\bibinfo{year}{2018}), \eprint{1807.01146}.

\bibitem[{\citenamefont{Czarnecki et~al.}(2004)\citenamefont{Czarnecki,
  Marciano, and Sirlin}}]{Czarnecki:2004cw}
\bibinfo{author}{\bibfnamefont{A.}~\bibnamefont{Czarnecki}},
  \bibinfo{author}{\bibfnamefont{W.~J.} \bibnamefont{Marciano}},
  \bibnamefont{and} \bibinfo{author}{\bibfnamefont{A.}~\bibnamefont{Sirlin}},
  \bibinfo{journal}{Phys. Rev.} \textbf{\bibinfo{volume}{D70}},
  \bibinfo{pages}{093006} (\bibinfo{year}{2004}), \eprint{hep-ph/0406324}.

\bibitem[{\citenamefont{Sirlin}(1967)}]{Sirlin:1967zza}
\bibinfo{author}{\bibfnamefont{A.}~\bibnamefont{Sirlin}},
  \bibinfo{journal}{Phys. Rev.} \textbf{\bibinfo{volume}{164}},
  \bibinfo{pages}{1767} (\bibinfo{year}{1967}).

\bibitem[{\citenamefont{Serebrov et~al.}(2018)}]{Serebrov:2017bzo}
\bibinfo{author}{\bibfnamefont{A.~P.} \bibnamefont{Serebrov}}
  \bibnamefont{et~al.}, \bibinfo{journal}{Phys. Rev.}
  \textbf{\bibinfo{volume}{C97}}, \bibinfo{pages}{055503}
  (\bibinfo{year}{2018}), \eprint{1712.05663}.

\bibitem[{\citenamefont{Pattie et~al.}(2018)}]{Pattie:2017vsj}
\bibinfo{author}{\bibfnamefont{R.~W.} \bibnamefont{Pattie}, \bibfnamefont{Jr.}}
  \bibnamefont{et~al.}, \bibinfo{journal}{Science}
  \textbf{\bibinfo{volume}{360}}, \bibinfo{pages}{627} (\bibinfo{year}{2018}),
  \eprint{1707.01817}.

\bibitem[{\citenamefont{Ezhov et~al.}(2018)}]{Ezhov:2014tna}
\bibinfo{author}{\bibfnamefont{V.~F.} \bibnamefont{Ezhov}}
  \bibnamefont{et~al.}, \bibinfo{journal}{JETP Lett.}
  \textbf{\bibinfo{volume}{107}}, \bibinfo{pages}{671} (\bibinfo{year}{2018}),
  \bibinfo{note}{[Pisma Zh. Eksp. Teor. Fiz.107,no.11,707(2018)]},
  \eprint{1412.7434}.

\bibitem[{\citenamefont{M{\"a}rkisch et~al.}(2019)}]{Markisch:2018ndu}
\bibinfo{author}{\bibfnamefont{B.}~\bibnamefont{M{\"a}rkisch}}
  \bibnamefont{et~al.}, \bibinfo{journal}{Phys. Rev. Lett.}
  \textbf{\bibinfo{volume}{122}}, \bibinfo{pages}{242501}
  (\bibinfo{year}{2019}), \eprint{1812.04666}.

\bibitem[{\citenamefont{Chang et~al.}(2018)}]{Chang:2018uxx}
\bibinfo{author}{\bibfnamefont{C.~C.} \bibnamefont{Chang}}
  \bibnamefont{et~al.}, \bibinfo{journal}{Nature}
  \textbf{\bibinfo{volume}{558}}, \bibinfo{pages}{91} (\bibinfo{year}{2018}),
  \eprint{1805.12130}.

\bibitem[{\citenamefont{Bazavov et~al.}(2019)}]{Bazavov:2018kjg}
\bibinfo{author}{\bibfnamefont{A.}~\bibnamefont{Bazavov}} \bibnamefont{et~al.}
  (\bibinfo{collaboration}{Fermilab Lattice, MILC}), \bibinfo{journal}{Phys.
  Rev.} \textbf{\bibinfo{volume}{D99}}, \bibinfo{pages}{114509}
  (\bibinfo{year}{2019}), \eprint{1809.02827}.

\bibitem[{\citenamefont{Czarnecki et~al.}(2020)\citenamefont{Czarnecki,
  Marciano, and Sirlin}}]{Czarnecki:2019iwz}
\bibinfo{author}{\bibfnamefont{A.}~\bibnamefont{Czarnecki}},
  \bibinfo{author}{\bibfnamefont{W.~J.} \bibnamefont{Marciano}},
  \bibnamefont{and} \bibinfo{author}{\bibfnamefont{A.}~\bibnamefont{Sirlin}},
  \bibinfo{journal}{Phys. Rev. D} \textbf{\bibinfo{volume}{101}},
  \bibinfo{pages}{091301} (\bibinfo{year}{2020}), \eprint{1911.04685}.

\bibitem[{\citenamefont{Kakazu et~al.}(2020)\citenamefont{Kakazu, Ishikawa,
  Ishizuka, Kuramashi, Nakamura, Namekawa, Taniguchi, Ukita, Yamazaki, and
  Yoshié}}]{Kakazu:2019ltq}
\bibinfo{author}{\bibfnamefont{J.}~\bibnamefont{Kakazu}},
  \bibinfo{author}{\bibfnamefont{K.-i.} \bibnamefont{Ishikawa}},
  \bibinfo{author}{\bibfnamefont{N.}~\bibnamefont{Ishizuka}},
  \bibinfo{author}{\bibfnamefont{Y.}~\bibnamefont{Kuramashi}},
  \bibinfo{author}{\bibfnamefont{Y.}~\bibnamefont{Nakamura}},
  \bibinfo{author}{\bibfnamefont{Y.}~\bibnamefont{Namekawa}},
  \bibinfo{author}{\bibfnamefont{Y.}~\bibnamefont{Taniguchi}},
  \bibinfo{author}{\bibfnamefont{N.}~\bibnamefont{Ukita}},
  \bibinfo{author}{\bibfnamefont{T.}~\bibnamefont{Yamazaki}}, \bibnamefont{and}
  \bibinfo{author}{\bibfnamefont{T.}~\bibnamefont{Yoshié}}
  (\bibinfo{collaboration}{PACS}), \bibinfo{journal}{Phys. Rev. D}
  \textbf{\bibinfo{volume}{101}}, \bibinfo{pages}{094504}
  (\bibinfo{year}{2020}), \eprint{1912.13127}.

\bibitem[{\citenamefont{Seng et~al.}(2020)\citenamefont{Seng, Galviz, and
  Mei{\ss}ner}}]{Seng:2019lxf}
\bibinfo{author}{\bibfnamefont{C.-Y.} \bibnamefont{Seng}},
  \bibinfo{author}{\bibfnamefont{D.}~\bibnamefont{Galviz}}, \bibnamefont{and}
  \bibinfo{author}{\bibfnamefont{U.-G.} \bibnamefont{Mei{\ss}ner}},
  \bibinfo{journal}{JHEP} \textbf{\bibinfo{volume}{02}}, \bibinfo{pages}{069}
  (\bibinfo{year}{2020}), \eprint{1910.13208}.

\end{thebibliography}

\end{document}